\documentclass[aps,prd,amsmath,amssymb,12pt]{revtex4}
\usepackage{graphicx}
\usepackage{bm}

\def\journal#1#2#3#4{{#1} {\bf #2} #3 (#4)}
\newcommand{\be}{\begin{equation}}
\newcommand{\ee}{\end{equation}}
\newcommand{\bea}{\begin{eqnarray}}
\newcommand{\eea}{\end{eqnarray}}
\newcommand{\hf}{\frac12}
\newcommand{\nn}{\cr}
\def\eq#1{(\ref{#1})}
\def\la{\langle}
\def\ra{\rangle}

\def\Tr{{\mathrm{Tr}}}
\def\ord#1{{\cal O}(#1)}
\def\mr#1{{\mathrm{#1}}}
\def\v#1{{\bm{#1}}}

\def\fd#1#2{\frac{\delta#1}{\delta#2}}
\def\fdd#1#2#3{\frac{\delta^2#1}{\delta#2\delta#3}}

\def\bx{\bar x}
\def\by{\bar y}
\def\bS{\overline S}
\def\bW{\overline W}
\def\tbj{{\bar{\tilde j}}}
\def\tbx{{\bar{\tilde x}}}
\def\tby{{\bar{\tilde y}}}
\def\tbz{{\bar{\tilde z}}}

\def\tj{{\tilde j}}
\def\tx{{\tilde x}}
\def\ty{{\tilde y}}
\def\tD{{\tilde D}}
\def\tG{{\tilde G}}
\def\tJ{{\tilde J}}
\def\ih{\frac{i}{\hbar}}
\def\sign{\mr{sign}}

\def\im{\mr{Im}}

\begin{document}
\title{Macroscopic limit of quantum systems}
\author{Janos Polonyi}
\affiliation{Strasbourg University, CNRS-IPHC,23 rue du Loess, BP28 67037 Strasbourg Cedex 2 France}
\begin{abstract}
Classical physics is approached from quantum mechanics in the macroscopic limit. The technical device to achieve this goal is the quantum version of the central limit theorem, derived for an observable at a given time and for the time-dependent expectation value of the coordinate. The emergence of the classical trajectory can be followed for the average of an observable over a large set of independent microscopical systems, and the deterministic classical laws can be recovered in all practical purposes, owing to the largeness of Avogadro's number. This result refers to the observed system without considering the measuring apparatus. The emergence of a classical trajectory is followed qualitatively in Wilson's cloud chamber.
\end{abstract}
\date{\today}
\maketitle

\section{Introduction}
The problem of recovering the laws of classical physics from the quantum level is beset by conceptual problems. The~exploration of the difference in the logical structure of the classical and quantum level was already started a long time ago by pointing out that the logic of quantum states is non-distributive as opposed to the Boolean logic of classical physics~\cite{birkhoff,mackey}. This structure was generalized later~\cite{svozil,chiara} but it remains to be seen how such an approach can help us to better understand the problem. The~solution may even lie beyond quantum theory and be related to the usual separation of physical laws and initial conditions~\cite{conway,hossenfelder}. A~slightly different trial starts with the construction of hidden parameter theories~\cite{belle} and seems to end at contextuality~\cite{bellec,kohen,mermin} without reconciliation of the two regimes in~sight. 

Maybe the most disturbing qualitative difference of the quantum and classical descriptions is the apparent non-deterministic nature of the choice of the spectrum element of an observable, which is realized in a measurement. One is advised to rely on a new dynamical mechanism {\cite{bassi}, such as spontaneous localization~\cite{ghirardir,ghirardip,smirne} or quantum stochastic processes~\cite{barchielli}, as a quick fix up of this selection process. In~the meantime, the phase transition, in~particular, the spontaneous symmetry breaking~\cite{anderson,neeman,grady,fioroni,pankovic,zimanyi,allahverdyan} has been proposed as an important ingredient of the collapse of the wave function, the~selection of the observed spectrum element, without~evoking new principles. A~more systematic approach to deal with logical incompatibilities is the decoherent histories program~\cite{griffithe,omnes,gellmann,griffithk}, which restricts attention only to the completely decohered sequences of~observations.

One can make a step toward a better understanding of these problems with the help of a peculiar feature of physical laws, namely, the measured physical values depend on the scales of the measurement. Furthermore, the quantitative dependence leads to the dependence of the physical laws on the scales of observation. For~instance, one extracts different physical laws when nature is observed with different spatial resolutions. Thus, the quantum--classical crossover can, in principle, be systematically discussed in the framework of the renormalization group, a~scheme to deal with the scale dependence of physical~laws. 

However, the scale dependence of physical laws is supposed to be smooth.} How can the logic and determinism change continuously between the quantum to the classical regimes? A hint to the answer might be found in more recent developments. The~need for increased accuracy of monitoring the motion of macroscopic bodies during the detection of gravitational waves has led to non-demolition measurements~\cite{braginsikyt},~weak measurements~\cite{braginsikyk},~a more careful treatment of quantum noise~\cite{clerk}, and to optomechanical devices~\cite{chen}. These examples show that classical concepts can be brought in agreement with quantum effects in a consistent and systematic manner. The~idea of continuous measurement~\cite{mensky} is another possibility to fit the quantum and the classical domains. The~common elements of these ideas is the introduction of a flexible length or time scale, which suppresses the uncertainties in the measurements, thereby approaching classical physics from the quantum level. Another development, which narrows the qualitative differences of the quantum and the classical level is the possibility that the quantum jump, assumed to make the choice of the measured value, is actually a dynamical process enfolding in finite time~\cite{ossiander,minev}. 

The proposition is put forward in the present work that the generalization of the central limit theorem to quantum mechanics provides a simple and generic framework to approach classical physics in the macroscopic limit. The~relation between the renormalization group and the central limit theorem was already discussed in the context of the generalized central limit theorem and the UV fixed points~\cite{jona,honkonen,koralov,calvo,amir}. The~central limit theorem is considered here from the point of view of quantum mechanics without the systematic methodology of the renormalization group. The~traditional classical laws are approached but never reached exactly, according to this scenario, in a manner similar to the way thermodynamics emerges from statistical physics. The~surprisingly large value of Avogadro's number renders the uncertainty of the average over a macroscopic sample negligible for all practical purposes without getting bogged down into the intricacy of contradicting logical structures. Such an approach of classical physics is not a new idea; it is the backbone of almost all attempts to arrive at the well-known classical domain from a microscopic level. The~modest steps are presented below, simply to draw attention to analogies of the way that the central limit theorem functions in classical probability~theory.

The macroscopic limit of quantum systems can be realized in different manners. The~typical rearrangement corresponds to the measurement of a microscopic property by a macroscopic measuring apparatus. To~comply with the probabilistic predictions of quantum mechanics, one considers a large set of equivalent and independent microscopic systems. Another possibility is to look for macroscopic quantum effects, where a single observed system of macroscopic size is used~\cite{leggett}. Such macroscopic quantum phenomena form a wide and colorful set. The~simplest is the indistinguishability of identical elementary particles, leading to the solution of Gibbs mixing paradox. It is perhaps the most universal macroscopic quantum effect, owing to its scale independence. Superconductivity or superfluidity are related to spontaneous symmetry breaking and are clear realizations of macroscopic phenomena, treated by an approximation valid only in the thermodynamical limit. In~general, the~semi-classical approximation can be used to discover a wide class of macroscopic quantum phenomena as saddle point effects~\cite{blasone}. The~non-unique definition of the actual number of degrees of freedom~\cite{shimizu,bjork,korsbakken,marquardt,frowis} and the non-triviality of their distribution over the possible states~\cite{kofler} indicate the richness of macroscopic quantum phenomena. A~common feature of the macroscopic limit in different systems is the decoherence of the macroscopic (collective) variables~\cite{joos,zureke,zurekk}. It is important to realize that the macroscopic variable may have an intrinsic environment within the system without a tensor product structure of the Hilbert space~\cite{saso}.  We restrict our attention in this work to the first case, where the measured phenomenon is microscopic and only the measuring apparatus amplifies it to a macroscopic~size. 

{We show that the average of an observable over a set of $N_s$ independent systems behaves classically up to $\ord{1/\sqrt{N_s}}$ quantum fluctuations and such a reduction of the quantum fluctuations takes place independently of the measurement process. The~actual measurement apparatus can be treated in a similar fashion to establish the almost classical nature of its macroscopic pointer variable. Hence, the measurement of an average over a large set of microscopic systems involves a correlation between two classical variables. However, the measurement of a single microscopic event reveals the usual difficulty of the measurement theory, the~dynamics of the amplification of a microscopic signal to a macroscopic level.}

We start with the extension of the central limit theorem over expectation values in Section~\ref{qclts} by calculating the second cumulant of the distribution of the average of an observable over of a set of $N_s$ independent microscopic systems. The~decrease of the width of the fluctuations is compatible with a reduction of the Planck constant $\hbar\to\hbar/N_s$. {The determination of the effective dynamics of the average over the microscopic system observables requires a scheme that can handle open interaction channels and remains valid at both sides of the quantum--classical crossover. Section~\ref{ctpfs} contains a brief motivation of the Closed Time Path (CTP) formalism used in this work.} The dynamical generalization of the central limit theorem for the average of the expectation value of a continuous coordinate is given in Section~\ref{coordinates}. The~role of the cumulants is taken over by the connected Green functions and they are shown to become suppressed as $N_s\to\infty$. The~result of the continuous observation is spoiled by the mass-shell singularities. To avoid this problem, the observations over a discrete set of time are considered. Furthermore, the condition of the applicability of the central limit theorem is given for an interactive set of microscopic systems. The~clarification of the role of a macroscopic measuring apparatus in the measuring process is taken up in Section~\ref{measurs}. It is shown that a harmonic measuring apparatus realizes a linear amplification, and Wilson's cloud chamber is briefly discussed, where the collapse of the wave function takes place as a dynamical process in a strongly coupled environment. Finally, the summary is presented in Section~\ref{summs}. Two appendices are included with some technical details. {The basic equations of the CTP propagator are collected} in Appendix \ref{ctps}. A~harmonic measuring apparatus is discussed in Appendix \ref{wcappa}.

\section{Central Limit Theorem for an~Observable}\label{qclts}
We consider the measurement of an observable $A$ on a closed microscopic system whose pure states belong to the Hilbert space $\cal H$. The~measurement is repeated on the ensemble of $N_s$ copies of the system; the~pure states of the ensemble form the direct product $\otimes{\cal H}^{N_s}$. The~{observable $A$ represented on} the $n$-th system, $A_n$ $n=1,\ldots,N_s$, acts on the $n$-th factor of the direct product space, and $\bar A=\sum_{n=1}^{N_s}A_n/N_s$ stands for the average observable { over the system ensemble}. If~the systems are indistinguishable, then one has to (anti-)symmetrize their state in $\otimes{\cal H}^{N_s}$ but this does not change the expectation value of $\bar A$ where the elements of the ensemble are not distinguished. We allow some variation of the {preparation of the systems within the ensemble, namely, we assume that $N_k=p_kN_s$ systems are placed into the state characterized by some reduced density matrix $\rho_k$ whose detailed specification is unnecessary at the moment; all we need is the normalization} $\sum_kp_k=1$.

The limit $N_s\to\infty$ can be realized by two different scenarios: (i) The textbook example consists of an ensemble of identically prepared, independent microscopical systems, accessed one-by-one, and~the expectation value of $\bar A$ is identified with the average of the measurements. (ii) Another, more realistic alternative is to consider a set of $N_s$ microscopic systems, where we measure directly $\bar A$, e.g., the center of mass of a mirror in a gravitation radiation detector or a total electric dipole moment of a solid in a linear response~study.

The central limit theorem can easily be recovered for transition amplitudes or expectation values. We discuss the latter; the~former can be recovered in similar manner. Let us start with the cumulant generator function of the observable $A$, defined by the following:
\be
e^{iw_k(j)}=\Tr[\rho_ke^{ijA}]=\la e^{ijA}\ra_k
\ee

We assume a continuous, unbound spectrum; the~extension of our procedure for a bounded or discrete spectrum is straightforward. The~probability distribution of $A$ is given by  the following Fourier transform:
\be
p_k(a)=\int\frac{dj}{2\pi}e^{-ija}\Tr[\rho_ke^{ijA}]
\ee
since {
\be
\la A^n\ra_k=\int dap_k(a)a^n.
\ee

It is easy to see that the generator function of $\bar A$, 
\be
e^{i\bar w(j)}=\la e^{ij\bar A}\ra=\sum_kp_k\la e^{ij\bar A}\ra_k,
\ee
}assumes  the following form:
\be
\bar w(j)=N_sw_q\left(\frac{j}{N_s}\right),
\ee
where $w_q(j)=\sum_kp_kw_k(j)$ is the average of the generator functions $w_k(j)$. One uses quenched averaging because the choice of the initial state influences the system for an arbitrarily long time in closed dynamics. It is a matter of a trivial expansion in $1/N_s$ to find the normal distribution around $\la A\ra$ of $\ord{1/\sqrt{N_s}}$ width,
\be
p(\bar a)=\sqrt{\frac{\la(A-\la A\ra^2)\ra}{2\pi N_s}}e^{-\frac{N_s}{2\la(A-\la A\ra^2)\ra}(\bar a-\la A\ra)^2}.
\ee

The decisive dynamical difference between scenarios (i) and (ii) is that while several measurements are performed in (i), only a single measurement is carried out in (ii). The~central limit theorem corresponds to scenario (i), and the dynamical aspect of the narrowing of the peak of $p(\bar a)$ in scenario (ii) is that the disturbance of a single measurement is distributed over a large number of systems. Such a point of view is consistent with the effective reduction of Plank's constant, $\hbar\to\hbar/N_s$, for~canonically conjugate pairs, $\bar p$ and $\bar q$. The~suppression of the incompatibility of canonically conjugate pairs opens the possibility of recovering a unique trajectory for the average observable $\bar A$. Hence the single measurement of scenario (ii) is non-demolishing when $N_s\to\infty$. The~mathematical equivalence of the two scenarios indicates that classical physics is reached by the averages in both~cases.
 
The non-demolishing nature of the measurement of $\bar A$ can simply be established in the Heisenberg representation. Let us assume that the ensemble of systems of scenario (i) and the measuring apparatus follow a closed dynamics with Hamiltonian $H=\sum_n(H_{s,n}+A_nB)+H_a$, where $H_a$ denotes the Hamiltonian of the apparatus, $H_{s,n}$ {and $A_nB$ represent the system Hamiltonian $H_s$ and the system-apparatus interaction energy $AB$, respectively, on the $n$-th microscopic system}, $B$ being an apparatus operator. The~Hamiltonian of scenario (ii) is the same, except $B=B'/N_s$. The~commutator $[\bar A,H]=[A,H_s]/N_s$ in the Schr\"odinger representation implies $[\bar A_H(t),H]=\ord{N_s^{-1}}$ in the Heisenberg representation {where $\bar A_H(t)=e^{\ih tH}\bar Ae^{-\ih tH}$} and the integration of the Heisenberg equation yields  the following:
\be\label{noninv}
[\bar A_H(t),\bar A_H(t')]=\ord{N_s^{-1}},
\ee
where the measurement is non-demolishing~\cite{braginsikyt} as $N_s\to\infty$. This result is formal as it stands since possible singularities in time or frequency are ignored; a~more careful discussion of this point follows~below.

The difference between the two scenarios is shown clearly by the dynamics of the measuring apparatus, namely, an apparatus observable, $A_a$, has the commutator $[A_a,H]=\ord{N^j_s}$ with $j=1$ and $0$ in the cases (i) and (ii), respectively. The~disturbance of the apparatus by the measurement increases with the number of measurements, and both the measured system and the measuring apparatus should be renewed after each~measurement.

{
\section{CTP~Formalism}\label{ctpfs}
The formalism to follow the macroscopical limit of quantum systems should be a CQCO scheme: it should handle classical (C) \cite{bateman,arrow,galley,kuwahara,perez} and quantum (Q) \cite{chou,jordan,calzetta,kamenev} dynamics because the macroscopical limit implies classical physics should further cover closed (C) and open (O) dynamics since a macroscopic system cannot be kept isolated on the microscopic scale. These possibilities are offered by the CTP formalism, initially proposed to deal with the perturbation expansion in the Heisenberg representation~\cite{schw} and with non-equilibrium phenomena in many-body systems~\cite{keldysh}. The~dynamics is defined in this work by a CTP action functional, used with the variational principle in classical mechanics and with the path integral in the quantum case. The~basic idea of this scheme is outlined below for a one-dimensional particle with coordinate $x$ within the time interval $t_i<t<t_f$.

\subsection{Classical Closed~Dynamics}
The motion is followed in the CTP scheme in both directions of time~\cite{irr} along the continuous trajectory $X(t)$, $t_i<t<2t_f-t_i$ in such a manner that the particle follows a second order equation of motion forward and backward in time for $t_i<t<t_f$ and $t_f<t<2t_f-t_i$, respectively. The~auxiliary conditions on the trajectories are $X(t_i)=X(2t_f-t_i)=x_i$, $\dot X(t_i)=-\dot X(2t_f-t_i)=v_i$, with~the given initial coordinate and velocity, $x_i$ and $v_i$, respectively. The~auxiliary conditions usually make the velocity discontinuous at $t_f$, at~the reversion of the time direction. The~trajectory $X(t)$ is then broken up into two trajectories by introducing $x_+(t)=X(t)$ and $x_-(t)=X(2t_f-t)$ for $t_i<t<t_f$. The~set of varied trajectories is, therefore, given by the pair of trajectories $\tx(t)=(x_+(t),x_-(t))$, which satisfy the initial conditions $x_\pm(t_i)=x_i$, $\dot x_\pm(t_i)=v_i$ and are closed at the final time, $x_+(t_f)=x_-(t_f)$ owing to the continuity of $X(t)$.

The action of the trajectory pair is the following:
\be\label{cctpact}
\tilde S[\tx]=S[x_+]-S[x_-]+i\frac\epsilon2\int_{t_i}^{t_f}dt[x^2_+(t)+x^2_-(t)],
\ee
where $S[x]$ denotes the traditional action. The~minus sign in front of the second term on the right hand side is induced by the reversed time direction for $x_-(t)$ and the infinitesimal imaginary term is added to split the degeneracy of the action, a~necessary condition to define Green functions, cf. Appendix \ref{ctps}.

The CTP generalization of the traditional action is optional in the case of closed dynamics, the~only advantage being the trading of the boundary conditions, required by the traditional variation principle, to~initial conditions. However, this change may be important for non-integrable~dynamics.

\subsection{Classical Open~Dynamics}
The traditional action principle of classical mechanics is applicable for systems with holonomic forces, represented by a term $L_h(x,\dot x)$ in the Lagrangian, which induces a conservative force, as follows:
\be
F_h=\frac{\partial L_h(x,\dot x)}{\partial x}-\frac{d}{dt}\frac{\partial L_h(x,\dot x)}{\partial\dot x},
\ee
in the Euler--Lagrange equation. This can be generalized to a semi-holonomic force~\cite{effth}, as follows:
\be\label{semiholf}
F_{sh}=\frac{\partial L_{sh}(x,x',\dot x,\dot x')}{\partial x}_{|x=x',\dot x=\dot x'}-\frac{d}{dt}\frac{\partial L_{sh}(x,x',\dot x,\dot x')}{\partial\dot x}_{|x=x',\dot x=\dot x'},
\ee
generated by a function $L_{sh}(x,x',\dot x,\dot x')$ in the Lagrangian. We separate here two roles of the trajectory $x(t)$, namely, the identification of the location of the particle and the variable on which the derivative acts. The~proper bookkeeping of such a separation is the reduplication of the degree of freedom, $x\to\tx$, introduced already above. The~imposition of the equation of motion and identical initial conditions for $x_+(t)$ and $x_-(t)$ sets $x_+(t)=x_-(t)$ and guarantees $x=x'$, $\dot x=\dot x'$ in \eq{semiholf}.

The simplicity of the action \eq{cctpact} such that the trajectories $x_\pm(t)$ are not coupled is a characteristic feature of closed dynamics. Let us now assume that the system and its environment together follow closed dynamics, given by the action $S[x,y]=S_s[x]+S_e[x,y]+i\frac\epsilon2\int dt[x^2(t)+y^2(t)]$ where $x$ and $y$ denote the system and the environment coordinate, respectively. It is easy to see that the elimination of the environment degrees of freedom by their equations of motion couples $x_+(t)$ and $x_-(t)$. The~form $\tilde S_{eff}[\tx]=S_1[x_+]-S^*_1[x_-]+S_2[x_+,x_-]$ of the action, obtained by separating the coupling between the trajectories in $S_2$, shows that the holonomic and the semi-holonomic forces are generated by $S_1$ and $S_2$, respectively. One can see furthermore that the open forces arising in any subsystem of closed dynamics are~semi-holonomic.

\subsection{Closed Quantum~Dynamics}
The reduplication of the degrees of freedom, needed to handle classical semi-holonomic forces by the variational principle, arises from quantum mechanics, as~well. In~fact, let us consider the expectation value 
\be\label{expval}
\la\psi|A|\psi\ra=\sum_{m,n}c_mc^*_n\la m|A|n\ra
\ee
of the observable $A$ in the pure normalized state $|\psi\ra=\sum_mc_m|m\ra$. This expression involves two representatives of the same state, a~bra and a ket, progressing in opposite directions in time. The~summation is over the quantum fluctuations in the basis $\{|m\ra\}$, and the factorizability of the coefficients $c_mc^*_n$ indicates that the quantum fluctuations in the bra and the ket are independent. The~independent dynamics of the bra and ket sector can be represented by employing two Hilbert spaces~\cite{umezawa}, giving rise to a reduplication of the physical operator set. This structure is optional for closed dynamics since the dynamics in the two Hilbert spaces are equivalent; they are related by Hermitian~conjugation.
 
The dynamical role of the reduplication is particularly clear when the expectation value \eq{expval} is considered at some time $t$, $\la\psi(t)|A|\psi(t)\ra=\Tr[A\rho(t)]$, where $\rho(t)=e^{-\ih tH}|\psi(0)\la\psi(0)|e^{\ih tH}$ stands for the density matrix. The~path integral expression for the density matrix, obtained by breaking up the two time evolution operators into the product of infinitesimal time propagation, is as follows:
\be\label{reddensm}
\la x_+|\rho(t_f)|x_-\ra=\int D[\tx]e^{\ih\tilde S[\tx]}\la x_+(t_i)|\rho(t_i)|x_-(t_i)\ra,
\ee
where one integrates over trajectories $x_\pm(t_f)=x_\pm$, and the action is given by \eq{cctpact}. The~two trajectories represent the independent quantum fluctuations in the bra and the ket~factors.

\subsection{Open Quantum~Dynamics}
The state of an open system is, in general, mixed, and its density matrix is not factorizable, as $\rho=\sum_np_n|n\ra\la n|$ with a sum of at least two contributions. Hence, the bra and the ket fluctuations are correlated and inseparable, and the formal reduplication of the degrees of freedom, mentioned for closed dynamics, becomes unavoidable. It is advantageous to use the Keldysh variables, namely, the average $x=(x_++x_-)/2$ and the difference $x_d=x_+-x_-$, representing the physical coordinate and the quantum fluctuations,~respectively.

The reduced system density matrix can be obtained by integrating out the environment degrees of freedom in the path integral, as follows:
\bea
\la x_+|\rho_s(t_f)|x_-\ra&=&\Tr_e[\rho(t_f)]\nn
&=&\int D[\tx]D[\ty]e^{\ih S[x_+,y_+]-\ih S^*[x_-,y_-]}\la x_+(t_i),y_+(t_i)|\rho(t_i)|x_-(t_i),y_-(t_i)\ra
\eea
where the integration is over the open system trajectories, $x_\pm(t_f)=x_\pm$, and~closed environment trajectories, $y_+(t_f)=y_-(t_f)$. The~effective system action $\tilde S[\tx]=S_s[x_+]-S_-^*[x_-]+S_{infl}[\tx]$ to be used in \eq{reddensm} for the reduced system density matrix now contains the influence action functional~\cite{feynman} given by the following:
\be
e^{\ih S_{infl}[\tx]}=\int D[\ty]e^{\ih S_e[x_+,y_+]-\ih S_e^*[x_-,y_-]}\la x_+(t_i),y_+(t_i)|\rho(t_i)|x_-(t_i),y_-(t_i)\ra.
\ee
The relation $\tilde S[x_+,x_-]=-\tilde S^*[x_-,x_+]$ follows from the unitarity of the underlying closed full~dynamics.

The CTP formalism offers a simple access to decoherence~\cite{joos,zureke,zurekk}. The~decoherence is a basis-dependent phenomenon, and its appearance in the coordinate representation can easily be seen by inspecting the path integral \eq{reddensm}. The~coordinate decoherence is qualitatively the suppression of the strongly off-diagonal elements of the density matrix $\rho(x_+,x_-)$. This is a dynamical process~\cite{dyndec} resulting from the system--environment interactions and can be recognized as a suppression of the contributions of trajectory  pairs $x_\pm(t)$ with large $x_d=x_+-x_-$. The~natural source of such a suppression is the increase of $\mr{Im}\tilde S[\tx]$ $x_d(t)$, and the strongly decohered semi-classical dynamics of a macroscopic variable corresponds to a path integral where only trajectory pairs $x_d(t)\approx0$ contribute. Such a path integral is actually a simple realization of the decoherent histories view of quantum~dynamics.

The dynamical origin of the reduplication of the degrees of freedom is similar in classical and quantum dynamics. The~reduplication is optional for closed dynamics and becomes a necessity to handle semi-holonomic forces and the mixed components of the state. The~reduplication leads to substantial complications in the equations, which may appear, at first sight, as unnecessary. However, these complications correspond to physical processes and their function is to enlarge the range of applicability of the~formalism.

The CTP scheme is used in this work to find the effective dynamics of the average coordinate over a ensemble of particles and of the pointer variable of the measuring apparatus. The~central limit theorem is easiest to state in terms of connected Green functions; it amounts to the claim that the limit $N_s\to\infty$ suppresses the $n$-th order Green function to $\ord{N_s^{n-1}}$. Few details about the second-order Green function, the~propagator, are collected in Appendix \ref{ctps}.
}

\section{Macroscopic Limit of the~Coordinate}\label{coordinates}
The observable averaged over a large number of independent microscopical systems becomes classical. What kind of effective dynamics do the average obey? The response to this question is sought by constructing the effective dynamics of the center of mass coordinate of a large set of non-interacting one-dimensional particles. {Another issue of the macroscopic limit of quantum systems is the emergence of decoherence, a~necessary condition of the classical limit. The~decoherence is a dynamical process; it is induced by the interaction with the environment, and~its understanding requires dynamical considerations.}

\subsection{Generator~Functional}
The dynamics of the coordinate $x$ of a particle is assumed to be defined by a CTP action $S_s[\tx]$. The~role of the cumulants is taken over by the connected Green functions whose generator functional is defined by the following path integral expression:
\be\label{octpspi}
e^{\ih W_\rho[\tj]}=\int D[\tx]e^{\ih S_\rho[\tx]+\ih\int dt\tj(t)\tx(t)},
\ee
where the action $S_\rho[\tx]=S_s[\tx]-i\hbar\ln\rho(x_+(x(t_i),x_-(x(t_i))$ contains the initial condition in terms of the initial density matrix $\rho$. The~parameterization $x_\pm=x\pm x_d/2$, $j^\pm=j/2\pm j_d$, $\tx\tj=x_+j_++j_-x_-=xj+x_dj_d$ is used below. The~knowledge of the generator functional $W_\rho[\tj]$ allows us to reconstruct $S_\rho[\tx]$ by functional Fourier transformation. We use the coordinate as our single observable; however, the extension of the subsequent arguments can easily be generalized over other observables by the use of an appropriately generalized generator~functional.

We now take $N_s$ independent copies of our system with initial density matrices belonging to the set $\{\rho_k\}$, where the state $\rho_k$ occurs $N_k=p_kN_s$ times. The~generator functional for the connected Green functions of the average coordinate, $\bx=\sum_{n=1}^{N_s}x_n/N_s$, is given by the following:
\be\label{cltqm}
\bW[\tj]=N_sW_q\left[\frac{\tj}{N_s}\right].
\ee
where $W_q[\tj]=\sum_kp_kW_{\rho_k}[\tj]$ is the quenched average of the generator functionals over different preparations. {The quantum central limit theorem states that the $n$-th order connected Green function of the average observable, defined by \eq{functexpgf}, is $\ord{N_s^{n-1}}$.}

\subsection{Continuous~Observations}\label{contobss}
The effective quantum action for $\bx$ can be obtained by the functional Fourier transformation, as follows:
\be\label{ftransf}
e^{\ih\bS_q[\tx]}=\int D[\tj]e^{\ih\bW[\tj]-\ih\int dt\tj(t)\tx(t)},
\ee
the case $N_s=1$ being trivial, $\bS_q[\tx]=S_s[\tx]$. When $N_s\gg1$, one finds the following:
\be\label{qclt}
\bS_q[\tx]=N_s\left[\hf\int dtdt'[\tx(t)-\tbx_{cl}(t)]\tD^{-1}(t,t')[\tx(t')-\tbx_{cl}(t')]+\ord{N_s^{-1}}\right]
\ee
where $\bx_{cl,\sigma}(t)=\delta W_q[0]/\delta j_\sigma(t)=\bx_{cl}(t)$ denotes the first moment of the averaged preparation and $D_{\sigma,\sigma'}(t,t')=-\delta^2W_q[0]/\delta j_\sigma(t)\delta j_{\sigma'}(t')$ stands for the propagator. The~factor $N_s$ can be interpreted as a simple rescaling of the Planck constant, $\hbar\to\hbar/N_s$. The~dynamics is seriously truncated in the macroscopic limit; the~first moment is kept untouched but the connected Green functions, representing the quantum fluctuations, are suppressed. The~difference between the classical physics of the non-fluctuating first moment as the trajectory and the quantum physics of a large but finite set of microscopic systems stems from the $\ord{N^{-1}_s}$ corrections.

The probability distribution of $\bx=(\bx_++\bx_-)/2$ can be obtained by setting $j_d=0$ and performing the Fourier transformation only in $j$, as follows:
\be\label{qcltpd}
p[\bx]=e^{\frac{N_s}{2\hbar}\int dtdt'[\bx(t)-\bx_{cl}(t)](D^i)^{-1}(t,t')[\bx(t')-\bx_{cl}(t')]+\ord{N_s^0}},
\ee
up to a normalization factor. The~positivity of the spectral density \eq{spectrd} makes $D^i(t,t')$ a negative definite operator. The probability distribution is Gaussian with an $\ord{N_s^{-1/2}}$ width around the expectation value. The~decoherence in the coordinate space, the~suppression of the contribution to the generator functional with increasing $\bx_d=\bx_+-\bx_-$, can be read off from the following:
\be
p_d(\bx_d)=e^{-\frac1\hbar\im\bS_q[\tbx]}
\ee
with
\be\label{decobs}
\im\bS_q[\tbx]=\frac{N_s}2\int dtdt'\bx_d(t)(\tD^{-1})^i(t,t')\bx_d(t')+\ord{N_s^0}.
\ee
Note that the kernel of the quadratic forms \eq{qcltpd} and \eq{decobs} differ.

Let us consider the time translation invariant limit $t_i\to-\infty$, $t_f\to\infty$ when the spectral function $\rho(\omega)=-D^i(\omega)/\pi$ for $\omega>0$ peaks at the quasi-particle frequencies. The~effective dynamics of $\bx$ is markedly different for the discrete and continuous spectra. In~the case of the discrete spectrum, $D^i$ displays Dirac-delta singularities at the stable particle modes, cf. the last equation in \eq{dnfi}. However, their contributions, the~on-shell stable particle modes, drop out from the harmonic action and their amplitudes acquire a uniform probability distribution, according to \eq{qcltpd}. The~dynamics remain coherent since, {according to the second equation of \eq{invprop},} $(D^{-1})^i(\omega)=\ord\epsilon$ at the quasi-particle peaks, cf. the last equation in \eq{hoingpro}. Such a state of affairs follows naturally from noting that the finite time observations can resolve the discrete spectral lines. The~spectrum of a macroscopic system is usually continuous and we cannot resolve the individual energy levels. { The quasi-particles, identified by long but finite time observations, are  narrow but non-trivial wave packets whose spread generates a finite lifetime. The~decay of such wave packets can be understood as the result of an interaction with the internal environment consisting of the other quasi-particle wave packets of finite time observations. It is this interaction which induces decoherence. The~spread of the spectral lines and the emergence of the decoherence can clearly be seen in Equation \eq{contsptrpr}, which corresponds to a harmonic dynamics with the continuous~spectrum. 

Equations \eq{qcltpd}--\eq{decobs} are the main results of this paper. Their lessons with finite $(\tD^{-1})^i$ kernel can be summarized as follows: The two CTP copies, representing the same physical degree of freedom, differ due to quantum fluctuations. The~latter are suppressed in the macroscopical limit and the unique classical trajectory is restored. Furthermore, the coefficient in front of the integral in \eq{decobs} indicates an $\ord{1/\sqrt{N_s}}$ decoherence length. Therefore, the decoherence becomes strong in the macroscopic limit, and the consistent history view of the quantum dynamics~\cite{griffithk} turns out to be an excellent approximation of the dynamics of a macroscopic collective coordinate. Finally, the~simple Gaussian suppression of the off-diagonal fluctuations represents a more systematical realization of the  phenomenological spontaneous localization scenario~\cite{ghirardir}.}

While the effective quantum dynamics for the average $\bx$ is given by the Fourier transformation \eq{ftransf}, the corresponding classical effective action is defined by the Legendre transformation as follows:
\be\label{cleffact}
\bS_{cl}[\tx]=\bW[\tj]-\int_t\tx(t)\tj(t),~~~\tx(t)=\fd{\bW[\tj]}{\tj(t)}.
\ee

In fact, the~inverse transformation, 
\be
\bW[\tj]=\bS_{cl}[\tx]+\int_t\tx(t)\tj(t),~~~-\tj(t)=\fd{\bS_{cl}[\tx]}{\tx(t)},
\ee
shows that the peak of the probability distribution \eq{qcltpd} satisfies the Euler--Lagrange equation of $\bS_x[\tx]$. The~quantum and the classical actions are usually different; however, the Fourier and the Legendre transformations agree for quadratic functions, leading to $\bS_{cl}[\tx]=\bS_q[\tx]+\ord{N_s^0}$. In~other words, Ehrenfest's theorem becomes valid since the width of a wave packet of $\bx$ is $\ord{N_s^{-1/2}}$. 

The action $\bS_q$ conveys a non-trivial message---the~suppression of the fluctuations. However, the simplicity of $\bS_{cl}$ is deceptive because the initial state is anchored in the action $S_\rho$ in the path integral \eq{octpspi}. An~equation of motion of classical physics is universal because the auxiliary conditions, needed to specify a given solution, are independent from them. {The~variational equation of $\bS_{cl}$, which is satisfied by the expectation value of the coordinate, is not universal} 
 since it already contains the initial conditions. We may, nevertheless, explore the physics of different average trajectories, $\bx_{cl}(t)$, by~using a suitable chosen  $N_s$-independent physical external source $j_{ph,\sigma}(t)=\sigma j_{ph}(t)$. The~procedure based on the external source $\sigma j_{ph}+j_\sigma/N_s$, $j_\sigma$ being the bookkeeping auxiliary variable used up to now, leads to a dynamics which is dominated by the trajectories in the $\ord{N_s^{-1/2}}$ vicinity of $\bx_{cl}(t)$.

\subsection{Discrete Observation~Times}
The formal result \eq{noninv} indicating the non-demolishing nature of the measurement for large $N_s$ seems to be supported by the $\ord{N_s^{-1/2}}$ width of the probability distribution \eq{qcltpd}. However, in closed dynamics, \eq{qcltpd} is actually flat for the normal modes defined by the discrete Dirac-delta peaks of the spectral~function. 

A less formal and more reliable argument for the emergence of a classical trajectory can be constructed with the help of a discrete set of measurements, carried out at discrete times, $t_\ell$, $\ell=1,\ldots,N_m$. The~external source is now written in the form $\tj(t)=\sum_\ell\tj_\ell\delta(t-t_\ell)$, and the generator functional $\bW[\tj]$ is reduced to a function of the coefficients $\tj_\ell$, as follows:
\be\label{cmgenf}
\bW_s(\tj)=\tj\tbx-\frac1{2N_s}\tj\tD\tj
\ee
where the quantities without time arguments are either $N_m$ dimensional vectors or the $N_m\times N_m$ matrix, written in component form as $\bx_{\sigma,\ell}=\delta W_q[0]/\delta j_\sigma(t_\ell)=\bx_\ell$ and $\tD_{\ell,\ell'}=-\delta^2W_q[0]/\delta\tj(t_\ell)\delta\tj(t_{\ell'})=\tD(t_\ell-t_{\ell'})$. The~corresponding action function,
\be\label{dqclt}
\bS_q(\tx)=\bS_{cl}(\tx)=-\frac{N_s}2(\tx-\tbx_{cl})\tD^{-1}(\tx-\tbx_{cl}),
\ee
differs from \eq{qclt} in that here, the inverse is that of a $2N_m\times2N_m$ matrix $\tD(t_\ell,t_{\ell'})$ as opposed to an operator $\tD(t,t')$ acting on time-dependent functions. The~physical origin of the difference is that while \eq{qclt} contains the full information about the harmonic dynamics, the \eq{dqclt} encodes a ``stroboscope physics'', the~dynamics at the time scales $t_j-t_{j-1}$ \cite{rgqm}. 

Let us consider first a single observation, whose result is summarized by the following distribution function:
\be
\chi(\tilde z_1)=\int d\tx e^{\ih\bS_q(\tx)}\delta(\tx_1-\tilde z_1).
\ee
The UV finiteness of the dynamics assures that the response to an external source builds up continuously in time and the retarded and the advanced propagators are vanishing when the external source is turned on or off, respectively, $D^r(0)=D^a(0)=0$. The~dependence on $j_{1d}$ is suppressed according to the block structure \eq{block}, and we find the following:
\be\label{distrsav}
\chi(\tilde z_1)=\delta(z_{1d})\sqrt{\frac{N_s\hbar}{-2\pi D^i(0)}}e^{\frac{N_s}{2D^i(0)\hbar}(z_1-\bx_1)^2}
\ee
with $D^i(0)<0$. The~observation leads to a collapse of the two members of the CTP doublet, a~complete decoherence, owing to the unitarity of the time evolution, and the common value follows a narrow Gaussian distribution, peaked at the expectation~value.

In the case of two observations, separated by the time $\tau=t_2-t_1>0$, the relevant distribution function,
\be
\chi(\tilde z_1,\tilde z_2)=\int d\tx_1d\tx_2 e^{\ih\bS_q(\tx_1,\tx_2)}\delta(\tx_1-\tilde z_1)\delta(\tx_2-\tilde z_2),
\ee
can be obtained by a straightforward integration, as follows:
\be\label{tdistr}
\chi(\tbz_1,\tbz_2)=\delta_{z_{d2},0}\sqrt{\frac{(2\pi N_s\hbar)^3}{-D^i(0)D^{r2}(\tau)}}e^{\frac{N_s}{2\hbar}\left[\frac{(z_1-\bx_1)^2}{D^i(0)}+2i\frac{z_{d1}[D^i(0)(z_2-\bx_2)-D^i(\tau)(z_1-\bx_1)]}{D^i(0)D^r(\tau)}+\frac{D^{i2}(0)-D^{i2}(\tau)}{D^i(0)D^{r2}(\tau)}z_{d1}^2\right]}.
\ee
The distribution of the result of the first measurement is identical to the case of a single measure, and the second, last measure is completely decohered. However,~the rest is difficult to understand, as there seems to be a non-trivial $z_{1d}$-dependence which influences the distribution of the result of the second measurement. Since the off-diagonality is not observable, it is better to integrate it out or equivalently, to~perform the Fourier transform of $\bW(\tj)$ for $j_d=0$, as follows:
\be
\chi(\bar z_1,\bar z_2)=\frac{2\pi\hbar N_s}{\sqrt{D^{i2}(0)-D^{i2}(\tau)}}e^{\frac{N_s}{4\hbar}\left[\frac{(\bar z^{(+)}-\bx^{(+)})^2}{D^i(0)+D^i(\tau)}+\frac{(\bar z^{(-)}-\bx^{(-)})^2}{D^i(0)-D^i(\tau)}\right]}
\ee
where $\bar z^{(\pm)}=\bar z_2\pm\bar z_1$ and  $\bx^{(\pm)}=\bx_{2,+}\pm\bx_{1,+}$. This is a more reasonable result, namely, the unitarity of the time evolution, the~non-observability of $x_d$, decouples the observations in the limit $N_s\to\infty$.

It is instructive to carry out the limit of two close observations, $\tau\to0$. The~distribution of the velocity $V=(\bx_2-\bx_1)/\tau$ and the coordinate $X=(\bx_1+\bx_2)/2$ for small $\tau\to0$ is as follows:
\be
\chi(X,V)=\frac{2\pi\hbar N_s}{\sqrt{-2D^{i}(0)\dot D^i(0)\tau}}e^{\frac{N_s}{2\hbar}\left[\frac{(X-x_{cl})^2}{D^i(0)}-\tau\frac{(V-\dot x_{cl})^2}{\dot D^i(0)}\right]}
\ee
with $\dot D^i(0)=dD^i(0)/dt$. The~spread of the distribution function in the velocity is consistent with the uncertainty principle with a reduced Planck constant, $\hbar\to\hbar/N_s$.

\subsection{Interactive Microscopic~Systems}
We have so far restricted our attention to non-interacting microscopic systems. It is natural to inquire about the condition on the strength of interactions among the systems which keep this result valid. To~find it, we consider an interactive family of microscopic systems with the generator functional for the single system Green functions as follows:
\be
e^{\ih W_s[\tj_1,\ldots,\tj_{N_s}]}=\int D[\tx]e^{\ih S_s[\tx]+\ih\sum_n\int dt\tj_n(t)\tx_n(t)},
\ee
where $S_s[\tx]=\sum_nS_{\rho_n}[\tx_n]$ denotes the action of the set of systems and $\rho_n$ stands for the density matrix of the initial state of the $n$-the system. The~generator functional can be written in the cluster expansion as the following:
\bea
W_s[\tj_1,\ldots,\tj_{N_s}]&=&\sum_{n_1=1}^{N_s}W_{s,n_1}[\tj_{n_1}]+\hf\sum_{n_1,n_2=1}^{N_s}W_{s,n_1,n_2}[\tj_{n_1},\tj_{n_2}]\nn
&&+\frac1{3!}\sum_{n_1,n_2,n_3=1}^{N_s}W_{s,n_1,n_2,n_3}[\tj_{n_1},\tj_{n_2},\tj_{n_3}]+\cdots
\eea
where the $\ell$-th term on the right hand side represents the $\ell$-body correlations. For~sufficiently weak coupled systems, the contributions of the correlations to the generator functional
\be
\bW[\tj]=W_s\left[\frac{\tj}{N_s},\ldots,\frac{\tj}{N_s}\right]
\ee
are suppressed as $N_s\to\infty$, 
\be\label{weakint}
\lim_{N_s\to\infty}\frac1{N_s^\ell}\sum_{n_1,\cdots,n_\ell=1}^{N_s}W_{s,n_1,\cdots,n_\ell}[\tj,\cdots,\tj]=0,
\ee
for $\ell\ge2$, and the systems can be considered independent. A sufficient condition of weakly coupled systems is that the each system is correlated with $\ord{N_s^0}$ others. This condition is satisfied by clusterizing Green functions displaying a finite correlation length in the thermodynamical limit and indicates a mean-field critical exponent that the fluctuations of the extensive quantities is growing with the square root of the~volume.

\section{Measurement~Process}\label{measurs}
We have so far looked into the emergence of a classical trajectory for the effective dynamics of the average over a large set of independent systems. However,~the average coordinate is measured by an apparatus, which is the subject of quantum mechanics as~well, and~it remains to be seen whether the process of the measurement preserves the classical nature of the average~coordinate.

\subsection{Measuring~Apparatus}\label{modaps}
We denote the coordinates of the degrees of freedom of the apparatus by $y_n$, \linebreak$n=1,\ldots,N_a$ and assume the form $S=S_s[\tx]+S_a[\ty]+\sum_{n=1}^{N_a}g_n\int dt\bx(t)y_n(t)$ for the action of the measured microscopic systems and the apparatus where the second and the third terms represent the action of the apparatus and the measured system--apparatus interaction, respectively, as well as $g_n=\ord{N_a^0}$. The~result of the measurement is read off from a pointer, a~macroscopic collective apparatus variable $\by=\sum_n\kappa_ny_n/N_a$ with $\kappa_n=\ord{N_a^0}$.

The generator functional for the apparatus Green functions,
\be
e^{\ih W_a[\tj_1,\ldots,\tj_{N_a}]}=\int D[\ty]e^{\ih S_a[\ty]+\ih\sum_n\int dt\tj_n(t)\ty_n(t)},
\ee
is written in  the following form:
\bea
W_a[\tj_1,\ldots,\tj_{N_a}]&=&\sum_{n_1=1}^{N_a}W_{a,n_1}[\tj_{n_1}]+\hf\sum_{n_1,n_2=1}^{N_a}W_{a,n_1,n_2}[\tj_{n_1},\tj_{n_2}]\nn
&&+\frac1{3!}\sum_{n_1,n_2,n_3=1}^{N_a}W_{a,n_1,n_2,n_3}[\tj_{n_1},\tj_{n_2},\tj_{n_3}]+\cdots
\eea
The apparatus is called weakly coupled if the correlations among the coordinates become suppressed as $N_a\to\infty$, 
\be\label{wcappc}
\lim_{N_a\to\infty}\frac1{N_a^\ell}\sum_{n_1,\cdots,n_\ell=1}^{N_a}W_{a,n_1,\cdots,n_\ell}\left[g_{n_1}\tilde\sigma\tbx_{cl}+\frac{\kappa_{n_1}}{N_a}\tJ,\ldots,g_{n_\ell}\tilde\sigma\tbx_{cl}+\frac{\kappa_{n_\ell}}{N_a}\tJ\right]=0,
\ee
for $\ell\ge2$. The~generator functional for the pointer Green functions is the following:
\be\label{pgenf}
e^{\ih W_P[\tJ]}=\int D[\tx]D[\ty]e^{\ih S_s[\tx]+\ih S_a[\ty]+\ih\sum_ng_n\int dt\tbx(t)\tilde\sigma\ty_n(t)+\ih\int dt\tJ(t)\tby(t)}
\ee
where the matrix $\tilde\sigma=\mr{Diag}(1,-1)$ takes care of the relative minus sign in the system--apparatus interaction term. The~indistinguishability of the measured systems can be taken into account by summing over the permutations $\pi\in S_{N_s}$ on the right hand side, by~calculating the path integral with the closing condition $x_{n,+}(t_f)=x_{\pi(n),-}(t_f)$ and by dividing the sum with $N_s!$. However, this is not necessary as long as we do not attempt to distinguish the $N_s$ systems by the measurement. The~systems may have a local interaction obeying the condition \eq{weakint}.

The integration over the microscopic system coordinates simplifies this expression to the following:
\be
e^{\ih W_P[\tbj]}=\int D[\ty]e^{\ih N_sW_q[\frac1{N_s}\tilde\sigma\tby]+\ih S_a[\ty]+\ih\int dt\tbj(t)\tby(t)}
\ee
and
\be\label{pathinfl}
W_P[\tbj]=W_a\left[g_1\tilde\sigma\tbx_{cl}+\frac{\kappa_1}{N_a}\tbj,\ldots,g_{N_a}\tilde\sigma\tbx_{cl}+\frac{\kappa_{N_a}}{N_a}\tbj\right]+\ord{N_s^{-1}}.
\ee

In the case of a weakly coupled apparatus, one has the following further simplification:
\be
W_P[\tJ]=\sum_{n=1}^{N_a}W_{a,n}\left[g_n\tilde\sigma\tbx_{cl}+\frac{\kappa_n}{N_a}\tJ\right],
\ee
leading to harmonic pointer effective action
\be\label{pointact}
S_Q[\tby]=\frac{N_a}2\int dtdt'[\tby(t)-\tby_{cl}(t)]\tG^{-1}_P(t,t')[\tby(t')-\tby_{cl}(t')]+\ord{N_a^0}
\ee
containing the pointer trajectory 
\be\label{ptraj}
\tby_{cl}(t)=\frac1{N_a}\sum_{n=1}^{N_a}\kappa_n\fd{W_{a,n}[\tj]}{\tj(t)}_{|\tj=g_n\tilde\sigma\tbx_{cl}}
\ee
and propagator
\be
\tG_P(t,t')=\frac1{N_a}\sum_{n=1}^{N_a}\kappa^2_n\fdd{W_{a,n}[\tj]}{\tj(t)}{\tj(t')}_{|\tj=g_n\tilde\sigma\tbx_{cl}},
\ee
cf. Equation \eq{qclt}. What we see here in the limit $N_s,N_a\to\infty$ is that a classical trajectory, $\tx_{cl}(t)$, generates another one, $\tby_{cl}(t)$. The~amplification of the microscopic $\tx_{cl}(t)$ to a macroscopic $\tby_{cl}(t)$ is linear in a harmonic apparatus, which is discussed in some details in \mbox{Appendix~\ref{wcappa}. }

The fluctuations may naturally play an important role in determining the relation between the $\tx_{cl}(t)$ and $\tby_{cl}(t)$. The~fluctuations within a macroscopic apparatus may become decisive according to \eq{ptraj} when some non-linearities generate unstable apparatus states, where the condition \eq{wcappc} is violated, the~typical example of non-clusterizing Green functions being a phase transition, cf. some early attempts to realize measurements in this manner~\cite{anderson,neeman,grady,fioroni,pankovic}. Such instabilities can be turned into an efficient non-linear amplifier~\cite{zimanyi,allahverdyan}; the~case of Wilson's cloud chamber is taken up briefly below. The~quantum fluctuations of the measured microscopic system may play an important role through the ignored $\ord{N_s^{-1}}$ contributions on the right hand side of \eq{pathinfl}. The~experimental demonstration of such quantum fluctuations is the goal of the interferometry of molecules~\cite{hornberger}.

\subsection{Particle Cloud~Chamber}
An example of a strongly coupled apparatus is Wilson's chamber~\cite{gupta,leone} containing a radioactive atom emitting, for example, an $\alpha$ particle. The~chamber is filled up with the mixture of air and water molecules, and its volume is suddenly extended to bring the mixture into supersaturation. The~ionization, caused by the $\alpha$-particle, triggers the condensation of water molecules into droplets along the straight classical~trajectory. 

The elementary units of the apparatus are the water molecules, and they should be handled by a statistical description. In~a stable thermodynamical phase, these molecules can be considered an ideal gas; however, the homogeneous unstable supersaturated state can approximately be taken into account by assuming a strong attractive interaction among the water molecules within the droplet size $r_{dr}$, supposed to be much larger than the wavelength of the $\alpha$-particle, $r_{dr}\gg\lambda_\alpha$. We have an apparatus that is weakly (strongly) coupled at distances $\ell\gg r_{dr}$ ($\ell\ll r_{dr}$). Two other important parameters of the chamber are used below: the~ratio of the water molecule density on the two sides of the first order phase transition, in~the droplet and in the supersaturated region, $n_{dr}/n_{ss}$, and~the average time needed for an ionization to take place, $\tau_i$. {The~action of the full system is the sum of the action \eq{cctpact} for the $\alpha$-particle with finite $t_i$ and $t_f$; an~action for the water molecules, $S_a[\v{y}_1,\ldots,\v{y}_{N_a}]$, and~the interaction term, assumed to be $g\int dtn(\v{x}_\alpha(t))$ where the density of water molecules , $n(\v{x})$, is taken along the $\alpha$-particle trajectory $\v{x}_\alpha(t)$.} 

We follow the ionization perturbatively and suppose that at order $N_m$, the ionizations which take place at the time and space, $t_j,\v{x}_j$, $j=1,\ldots,N_m$, are separated spatially more than the droplet size, $|\v{x}_{j+1}-\v{x}_j|>r_{dr}$, an~assumption to be justified later. By~neglecting the interactions among the droplets, the contribution to the probability of propagation of the $\alpha$-particle from $\v{x}_i=\v{x}_0=\v{x}_{0,\pm}$ at $t=t_i=t_0$ to $\v{x}_f=\v{x}_{N_m+1}=\v{x}_{N_m+1,\pm}$ at $t=t_f=t_{N_m+1}$ is as follows:
\be\label{perttraj}
\prod_{j=1}^\ell\int d^3\tx_j\prod_{j'=0}^\ell D_0(\v{x}_{j'+1,+}-\v{x}_{j',+},t_{j'+1}-t_{j'})\prod_{j''=0}^\ell D_0^*(\v{x}_{j''+1,-}-\v{x}_{j'',-},t_{j'+1}-t_{j'})
\ee
up to a trajectory independent constant where 
\be
D_0(\v{x},t)=\left(\frac{m}{2\pi i\hbar t}\right)^\frac32e^{\frac{im}{2\hbar t}\v{x}^2}
\ee
denotes the free propagator. This is a Gaussian integral and can simply be calculated up to a time-dependent constant by evaluating the integrand at the saddle point. It is easy to show by recursion in $N_m$ that the saddle point is a straight line, $\v{x}_j=(t_j-t_0)(\v{x}_{\ell+1}-\v{x}_0)/(t_{\ell+1}-t_0)+\v{x_0}$ in agreement with the second-order perturbation expansion in the operator formalism~\cite{mott,antonio}.

A pointer variable, the~center of mass of the corresponding droplet, is formed dynamically at each each ionization. The~number of elementary constituents, the~water molecules, in~a droplet, is large enough to render the pointer classical. In~a good approximation, the ionization depletes the water density in a sphere of radius $r_{dr}(n_{dr}/n_{ss})^{1/3}$. The~next droplet should be formed at a place where there are sufficient water molecules around, justifying $|\v{x}_{j+1}-\v{x}_j|>r_{dr}$. Hence, the apparatus can be considered to be weakly coupled in the selection of the ionization location, and the definition of the pointer value is guided by the integrand of \eq{perttraj}. 

We have a single $\alpha$-particle, $N_s=1$; hence, one expects large fluctuations around the saddle point, which in turn renders the value of the pointer almost random. However a closer look at the integrand of \eq{perttraj} reveals that this is not the case. In~fact, let us evaluate the integral by a coarse graining in two steps: We divide the volume of the chamber into small cubes of size $r_{dr}$ and integrate first within the cubes and after that sum over them. One finds that the assumption $r_{dr}\gg\lambda_\alpha$ leads to a strong suppression of the integral within a cube, owing to a fast phase oscillation unless the centers of the cubes correspond to the motion of a free particle. Therefore, the chain of droplets of a single $\alpha$-particle lies on a straight line within the resolution of the size of the~droplets.

The strong short range attractive force among the water molecules plays several roles. It drives the amplification, mentioned at the end of Section~\ref{modaps}, in~such a peculiar manner that the classical trajectory of the pointer(s) is formed in a weakly coupled apparatus and the homogeneous integral measure in \eq{perttraj} is justified. In~addition, it is responsible for the suppression of quantum fluctuations at $N_s=1$ and places the chain of ionizations along a classical trajectory. Finally, it regenerates the pointer variable in a dynamical manner for each measurement, i.e., the~droplet. This mechanism is a dynamical realization of the hypothetical step, the~type 1 sudden quantum state change of Neumann~\cite{neumann} or alias collapse of the wave~function.

Which classical trajectory is chosen where the wave function collapses to? Let us follow the propagation of the $\alpha$-particle from $t=t_i$ at $\v{x}=\v{x}_i$. The~time and location of the first ionization is chosen randomly~\cite{schonfeld}. This step, the~core of the measurement problem, remains unsolved, but as soon as this step is made, the rest follows from a semi-classical picture: The {emergence of the first droplet breaks the initial rotational symmetry into an axial symmetry} consisting of rotations around $\v{x}_1-\v{x}_0$. This symmetry is approximately preserved by the successive ionizations, which line up with an average separation $\tau_i(\v{x}_1-\v{x}_0)/(t_1-t_0)$ along a straight~line.

\section{Summary}\label{summs}
The proposal is put forward in this work that one can avoid the logical conflict between quantum and classical physics when the latter is considered to be the macroscopic limit of the former. The~quantum generalization of the central limit theorem is given by showing the narrowing of the probability distribution of the average observable over an $N_s$ independent identical system in the limit $N_s\to\infty$. {A strong Gaussian suppression of the off-diagonal fluctuations is found, which removes the difference between the two CTP copies of the same physical degree of freedom, thereby restoring the unique classical trajectory, in agreement with the consistent history view of the macroscopic limit and the idea of the spontaneous localization}. Therefore, the proposition is to regard classical physics as the limit of large but finite quantum systems, where the size of Avogadro's number suppresses quantum fluctuations for all practical purposes.  The~conditions on the strength of the correlations among the microscopic systems are given to preserve the simplicity of the result. The~observation over a discrete set of times reveals the discrete dynamics, characteristic at the time scale of the~observations. 

These results correspond to the effective dynamics of an observable, without~considering the measurement---the~process of the observation. When the central limit theorem is applied to the pointer variable of the measuring apparatus, it is advantageous to separate the case of measuring an average or a single quantity. The~measurement of an average observable over a large ensemble of microscopic systems implies a classical correlation between two classical trajectories: that of the average measurement and the pointer. A~harmonic apparatus yields a linear amplification of the measured signal within the realm of the central limit theorem. To~achieve a non-linear amplification, we need correlations among the constituents of the apparatus beyond the limit the applicability of the quantum central limit. In~the case of a single microscopic measurement, the linear amplification is obviously insufficient, and the measuring apparatus must be strongly correlated; the line of thought followed here is not applicable.

It is argued that Wilson's cloud chamber represents a simple set of both weakly and strongly correlated constituents, where on the one hand, one can use the central limit theorem for the pointer variable and, on the other hand one, can retain the quantum fluctuations of the ionizing particle. The~droplet formation in the chamber is a dynamical realization of the assumed collapse of the wave function. This view suggests that Neumann's postulate of a sudden change of the quantum state can be avoided, and the impulsive nature of the quantum jump results from a large number in the dynamics, $N_s$, which generates new time~scales. 

This idea is a natural extension of the view of the measurement as a spontaneous symmetry breaking, the~breakdown of the initial rotational symmetry by the formation of the first droplet. The~high degeneracy of the symmetrical ground state before the measurement is the dynamical origin of the strong correlation of the microscopic system and the measuring apparatus. In~fact, the~initial microscopic inhomogeneities of the gas exert a strong impact on the probability distribution of the location of the first droplet when the gas dynamics is taken into account by the (almost) degenerate perturbation~expansion. 

However, any further argument within the framework of quantum mechanics or quantum field theory can only push back the origin of the wave function collapse to earlier initial conditions without providing a deterministic origin. Therefore, the main problem of measurement theory, the~choice of the observed spectrum value of an observable, remains untouched by these~considerations.

The central limit theorem, used in this work, relies on a set of independent or weakly coupled microscopic systems. One may continue this line of thought and reach more realistic models by moving away from the vicinity of a Gaussian fixed point. This step implies the quantum  renormalization group~\cite{qrg} requiring the use of quantum field theoretical~methods.

\appendix
\section{CTP~Propagator}\label{ctps}
{This appendix is to introduce the Green functions and to present some important properties of the second order Green function: the~propagator.

\subsection{Generator~Functional}
The hierarchy of Green functions is easiest to define by means of generator functionals. In~the classical case, this functional for the connected Green functions is as follows:
\be
W[\tj]=\tilde S[\tx]+\int_{t_i}^{t_f}dt\tj(t)\tx{t}
\ee
where the trajectory satisfies the equation of motion $\delta\tilde S[\tx]/\tx(t)=-\tj(t)$. The~unique definition of this functional requires that the action be non-degenerate. The~real part of the action \eq{cctpact} displays two kinds of degeneracy. The~real part of the action is vanishing for physically realizable classical trajectories; $x_+(t)=x_-(t)$ and the normal modes of a quadratic action drop out. Both degeneracies are cured by the infinitesimal imaginary part.} The bra and the ket dynamics within the time interval $t_i<t<t_f$ can be captured by the generator functional of the connected Green functions for the following coordinate:
\be\label{pgenfunctop}
e^{\ih W[\tj]}=\Tr[U[t_f,t_i;j_+]\rho(t_i)U^\dagger[t_f,t_i;-j_-]]
\ee
where $U[t_f,t_i;j]$ denotes the time evolution operator for the time interval $(t_i,t_f)$ with $t_i\le t_m\le t_f$ in the presence of a time-dependent external source in the Hamiltonian, $H\to H-jx$, and~$\rho(t_i)$ stands for the initial density matrix. The~generating functional is an analytic functional for stable dynamics and the connected Green functions are defined by the functional Taylor expansion, as follows:
\be\label{functexpgf}
W[\tj]=\sum_{n=0}^\infty\frac1{n!}\sum_{\sigma_1,\ldots,\sigma_n}\int dt_1\cdots dt_nj_{\sigma_1}(t_1)\cdots j_{\sigma_1}(t_1)D_{\sigma_1,\ldots,\sigma_n}(t_1,\ldots,t_n)
\ee
with $\tj=(j_+,j_-)$. The~path integral representation of this generator function is the following:
\be\label{genfunc}
e^{\ih W[\tj]}=\int D[\tx]e^{\ih\tilde S[\tx]+\ih\int dt\tj(t)\tx(t)}
\ee
where the action is given by \eq{cctpact}, the~imaginary part assuring the convergence of the path integral. The~integration is over the trajectory pairs $\tx_\pm(t)=(x_+(t),x_-(t))$, $t_i<t<t_f$, satisfying the closing condition $x_+(t_f)=x_-(t_f)$, which represents the trace operation in \eq{pgenfunctop}, and the convolution with the initial state is suppressed for easier readability. The~closing of the trajectories couples $x_+$ and $x_-$, breaks the translation invariance in time and violates the energy conservation. To~eliminate such formal problems, we take the limit $t_f\to\infty$. This limit is non-trivial~\cite{irr}; however, a simple way to find it for a harmonic oscillator is shown in Appendix~\ref{props}, which trivially generalizes for interactive~systems.

The unitarity of the closed dynamics of the full observed system and its environment imposes important constraints on the generator functional $W[\tj]$. The~unitarity holds for an arbitrary physically realizable external source, $j_\pm=\pm j$; hence, the conservation of the probability implies $W[\tj]_{|j_+=-j_-}=0$. A~surprising corollary is that all Green functions of $x_d=x_+-x_-$ are vanishing, and the quantum fluctuations, represented by $x_d$, are ``invisible'' alone, without~coupling them to observables. Another important result of the unitarity is the invariance of the generator functional under the change the final time $t_f$ in the presence of physical source, $j_\pm=\pm j$.

\subsection{Propagator}\label{props}
The propagator
\be
D_{\sigma,\sigma'}(t,t')=-i\hbar\begin{pmatrix}\la T[x(t)x(t')]\ra&\la x(t')x(t)\ra\cr\la x(t)x(t')\ra&\la T[x(t)x(t')]\ra^*\end{pmatrix}
\ee
is symmetric, $D_{\sigma',\sigma'}(t,t')=D_{\sigma',\sigma}(t',t)$, and~the identity $T[AB]+(T[AB])^\dagger=AB+BA$ implies $D_{++}+D_{--}=D_{+-}+D_{-+}$. The~physical source $j_\pm=\pm j$ generates the expectation value as follows:
\be
\la x_\pm(t)\ra=\int_{-\infty}^\infty dt[D_{\pm,+}(t,t')-D_{\pm,-}(t,t')]j(t')
\ee
which is real and independent of the choice of the sign $\pm$. These constraints allow us to represent the propagator in terms of three real functions, $D^n(t,t')=D^n(t',t)$, $D^i(t,t')=D^i(t',t)$ and $D^f(t,t')=-D^f(t',t)$, as follows:
\be\label{block}
\tD=\begin{pmatrix}D^n+iD^i&-D^f+iD^i\cr D^f+iD^i&-D^n+iD^i\end{pmatrix}.
\ee
The components $D^n$  and $D^f$ are called near and far Green's function, by analogy with classical~electrodynamics. 

The propagator is easy to find in the operator formalism for a harmonic oscillator of mass $m$ and frequency $\omega_0$, as follows:
\be\label{frprop}
\tD(t,t')=-\frac{i}{2m\omega_0}\begin{pmatrix}e^{-i|t-t'|\omega_0}&e^{i(t-t')\omega_0}\cr e^{-i(t-t')\omega_0}&e^{i|t'-t|\omega_0}\end{pmatrix}.
\ee
Its expression in the frequency space is the following:
\bea\label{hogrfnct}
\tD_{\omega_0}(\omega)&=&\int dte^{i\omega(t-t')}\tD_{\omega_0}(t-t')\nn
&=&\frac1m\sum_n\begin{pmatrix}\frac1{\omega^2-\omega^2_0+i\epsilon}&
-2\pi i\delta(\omega^2-\omega^2_n)\Theta(-\omega)\cr-2\pi i\delta(\omega^2-\omega^2_n)\Theta(\omega)&-\frac1{\omega^2-\omega^2_0-i\epsilon}\end{pmatrix},
\eea
or
\bea\label{dnfi}
D^n_{\omega_0}(\omega)&=&P\frac1{m(\omega^2-\omega_0^2)},\nn
D^f_{\omega_0}(\omega)&=&-i\frac{\pi}m\delta(\omega^2-\omega_0^2)\sign(\omega),\nn
D^i_{\omega_0}(\omega)&=&-\frac{\pi}m\delta(\omega^2-\omega_0^2),
\eea
where $P$ denotes the principal part. The~limit $t_i\to-\infty$, $t_f\to\infty$ is trivial to carry out in \mbox{\eq{frprop}} yielding the action of a harmonic oscillator,
\be
\tilde S_{HO}=\hf\int_{-\infty}^\infty dtdt'\tx(t)\tD^{-1}(t-t')\tx(t').
\ee

{Two remarks are in order at this point. First, one finds the same second order Green function for the classical and the quantum harmonic oscillator. The~second concerns the generalization of the limit $t_i\to-\infty$, $t_f\to\infty$ for the interactive case, where one can simply retain the imaginary terms corresponding to the quadratic part of the action as in the case of a harmonic oscillator.}

\subsection{Inverse~Propagator}\label{invprops}
The propagator \eq{frprop} displays time translation invariance; hence, $D^n$, $D^f$ and $D^i$ are commutative, leading to the inverse with the block structure as follows:
\be\label{iblock}
\tD^{-1}=\begin{pmatrix}(\tD^{-1})^n+i(\tD^{-1})^i&(\tD^{-1})^f-i(\tD^{-1})^i\cr-(\tD^{-1})^f-i(\tD^{-1})^i&-(\tD^{-1})^n+i(\tD^{-1})^i\end{pmatrix}
\ee
where
\bea\label{invprop}
(\tD^{-1})^{\stackrel{r}{a}}&=&(\tD^{-1})^n\pm(\tD^{-1})^f=(D^{\stackrel{r}{a}})^{-1},\nn
(\tD^{-1})^i&=&-(D^r)^{-1}D^i(D^a)^{-1}.
\eea
In particular, the~kernel of the harmonic oscillator action,
\be
\tD^{-1}_{\omega_0}(\omega)=m\left[(\omega^2-\omega_0^2)\begin{pmatrix}1&0\cr0&-1\end{pmatrix}
+i\epsilon\begin{pmatrix}1&-2\Theta(-\omega)\cr-2\Theta(\omega)&1\end{pmatrix}\right],
\ee
gives  the following:
\bea\label{hoingpro}
(D_{\omega_0}^{-1})^n&=&m(\omega^2-\omega_0^2),\nn
(D_{\omega_0}^{-1})^f&=&im\sign(\omega)\epsilon,\nn
(D_{\omega_0}^{-1})^i&=&m\epsilon,
\eea
when the Lorentzian Dirac-delta $\delta(\omega)=\epsilon/\pi(\omega^2+\epsilon^2)$ is used in the propagator. The~CTP action for a closed dynamics, defined by the traditional action $S[x]$ is, therefore, the following:
\be
\tilde S[\tx]=S_s[x_+]-S_s[x_-]+\frac\epsilon\pi\int dtdt'\frac{x^-(t)x^+(t')}{t-t'+i\epsilon}+\frac{i\epsilon}2\int dt[x^{+2}(t)+x^{-2}(t)].
\ee
The comparison with Equation \eq{cctpact} reveals that the coupling of the trajectories at the final time is replaced in the limit $t_f\to\infty$ by an infinitesimal, time translation invariant coupling over the whole time~evolution. 

The block structure of the inverse propagator is similar to $\tilde\sigma\tD\tilde\sigma$ where $\tilde\sigma=\mr{Diag}(1,-1)$ is the ``metric tensor'' of the CTP formalism. The~calculation of the propagator, given by the Schwinger--Dyson resumed form, as in Equation \eq{pontprop}, is facilitated by noting that the multiplication of the block matrices 
\be
\hat A_\sigma=\sigma\hat A=\begin{pmatrix}A^n+iA^i&-A^f+iA^i\cr-A^f-iA^i&A^n-iA^i\end{pmatrix}
\ee
yields  the following relations:
\bea\label{prodas}
(A_{\sigma1}\cdots A_{\sigma n})^{\stackrel{r}{a}}&=&A_{\sigma1}^{\stackrel{r}{a}}\cdots A_{\sigma n}^{\stackrel{r}{a}}\nn
(A_{\sigma1}\cdots A_{\sigma n})^i&=&\sum_{j=0}^n(A_{\sigma 1}\cdots A_{\sigma j})^rA_{\sigma j+1}^i(A_{\sigma j+2}\cdots A_{\sigma n})^a.
\eea

\subsection{Continuous~Spectrum}
While the spectrum of finite systems is usually discrete, the density of state becomes so high for macroscopic systems that the resolution of the elementary spectrum lines is impossible within the time available for observations. Hence, models with a continuous spectrum are better suited to describe the macroscopic~limit. 

The continuous spectrum version of our harmonic model for the measuring apparatus is defined by spectral density 
\be\label{spectrd}
\rho(\omega)=\frac{i}{2\pi}D_{-+}(\omega)
\ee
which is vanishing for $\omega\le0$ and non-negative for $\omega\ge0$, $D^f(\omega)=\sign(\omega)iD^i(\omega)$, when the initial state is the ground state. The~propagator can then be brought into  the following form:
\be
\tD(\omega)=2\int_0^\infty d\Omega\Omega\rho(\Omega)\tD_\Omega(\omega)
\ee
assuming that the normal modes are rescaled to have the same mass, $m$.

As of a phenomenologically motivated parameterization, the~Drude-like $\ord{\Omega^{-1}}$ spectral function produces divergent Green functions for $t=0$. Hence, it is better to use the following:
\be
\rho(\Omega)=\Theta(\Omega)\frac{\lambda^2\Omega\Lambda}{(\Lambda^2+\Omega^2)^2},
\ee
giving
\bea\label{contsptrpr}
D^r(\omega)&=&\frac{\lambda^2\pi}{2m(\omega+i\Lambda)^2},\nn
D^i(\omega)&=&-\frac{\lambda^2\pi|\omega|\Lambda}{m(\omega^2+\Lambda^2)^2}\nn
(D^{-1})^r(\omega)&=&\frac{2m}{\lambda^2\pi}(\omega+i\Lambda)^2,\nn
(D^{-1})^i(\omega)&=&\frac{4m\Lambda|\omega|}{\lambda^2\pi},
\eea
in particular,
\bea
D^r(t)&=&-\Theta(t)\frac{t\lambda^2\pi e^{-\Lambda t}}{2m},\nn
D^i(t)&=&-\frac{\lambda^2}{\pi m\Lambda}\int_0^\infty dz\frac{z\cos z\Lambda t}{(z^2+1)^2},
\eea
and $D^i(0)=-\lambda^2/2\pi m\Lambda$.

\section{Gaussian Measuring~Apparatus}\label{wcappa}
We assume that the measuring apparatus of Section~\ref{modaps} follows harmonic dynamics with normal coordinates $y_n$, $n=1,\ldots,N_a$, obeying  the following action:
\be 
S_a[\ty]=\hf\sum_{n=1}^{N_a}\int dtdt'\ty_n(t)\tG_n^{-1}(t,t')\ty_n(t'),
\ee
with time translation invariance, $\tG(t,t')=\tG(t-t')$.

\subsection{Continuous~Measurement}
The pointer action \eq{pointact} now contains  the following trajectory:
\be\label{contpointtr}
\tby_{cl,\sigma}(t)=\frac1{N_a}\sum_n\kappa_ng_n\int dt'G^r_{D,n}(t,t')\bx_{cl}(t')
\ee
and the propagator
\be\label{pontprop}
\tG_P(t,t')=\frac1{N_a}\sum_n\kappa^2_n\tG_{D,n}(t,t'),
\ee
where
\be
\tG_{D,n}(t,t')=\left[\tG_n^{-1}+\frac{g_n}{N_s}\tilde\sigma\tD\tilde\sigma\right]^{-1}(t,t')
\ee
is the $n$-th normal mode propagator given in a form reminiscent of a Schwinger--Dyson resumed expression, the~second term in the denominators being the self energy. It is a characteristic feature of the harmonic dynamics that the self energy remains a simple one-loop expression, the~loop summation running over the index $n$. The~inverse propagator in the denominator is usually an unbounded operator, but the expansion in the self energy is possible and leads to a geometrical series, showing that the pointer excitations spend part of their lifetime in the measured system. This contribution is $\ord{N_s^{-1}}$;  the~impact of the pointer--microscopic system interaction on the apparatus propagator is suppressed in the thermodynamical~limit.

The inverse pointer propagator is given by $(G_P^{-1})^{\stackrel{r}{a}}=(G_P^{\stackrel{r}{a}})^{-1}$ and the following:
\be
(G_P^{-1})^i(t,t')=-\int dt_1dt_2(G_P^r)^{-1}(t,t_1)G_P^i(t_1,t_2)(G_P^a)^{-1}(t_2,t')
\ee
in terms of $\tG_P$. The~probability distribution of the pointer trajectories is given by the following:
\be\label{realef}
p[\by]=e^{-\frac{N_a}2\int dtdt'[\by(t)-\by_{cl}(t)](G_P^i)^{-1}(t,t')[\by(t')-\by_{cl}(t')]}
\ee
up to a normalization factor, and the decoherence is displayed by the following:
\be\label{imagef}
\mr{Im}S_P[\tby]=\frac{N_a}2\int dtdt'ty_d(t)(G_P^{-1})^i(t,t')ty_d(t'),
\ee
cf. \eq{qcltpd} and \eq{decobs}. The~first moments, the~expectation values of the coordinates, satisfy the classical equation of motion in harmonic dynamics; therefore, a harmonic apparatus shows no genuine quantum effects and serves only as a linear~amplifier.

\subsection{Discrete Sequence of~Measurements}
To decouple the mass-shell singularities of the apparatus, we consider measurements performed at a discrete time series, $t_\ell$ $\ell=1,\ldots,N_m$, and~described by time-dependent coupling constants, $g_n(t)=g_n\sum_\ell\delta(t-t_\ell)$. Note that despite the discrete set of the measurement times, the pointer is supposed to be followed continuously in time. The~influence of the microscopic systems on the apparatus is encoded by the distribution of the average coordinate at the measurement times, as follows:
\be
\chi(\tx)=\delta(x_{N_m,d})e^{-i\frac{N_s}{2\hbar}(\tx-\tbx_{cl})\tD^{-1}(\tx-\tbx_{cl})}.
\ee
It is important to bear in  mind that the index $\ell$ is suppressed in this, and the following expressions and the inverse matrix in the exponent are defined by $\sum_{\sigma'',\ell''}D_{(\sigma,\ell),(\sigma'',\ell'')}D^{-1}_{(\sigma'',\ell''),(\sigma',\ell')}=\delta_{\sigma,\sigma'}\delta_{\ell,\ell'}$. i.e.,~$\tD^{-1}_{\ell,\ell'}\ne\tD^{-1}(t_\ell,t_{\ell'})$. The~unitary of the time evolution imposes $\bx_{d,N_m}=0$ at the last~measurement. 

The generator functional for the connected pointer Green functions,
\be\label{genfpond}
e^{\ih W[\tJ]}=\int d\tbx\chi(\tbx)\int D[\ty]e^{\frac{i}{2\hbar}\sum_n\int dtdt'\ty_n(t)\tG^{-1}_n(t,t')\ty_n(t')+\ih\int dt\tJ(t)\tby(t)+\ih\sum_ng_n\int dt\tbx(t)\tilde\sigma\ty_n(t)},
\ee
leads after performing some Gaussian integrals to the pointer quantum action,
\be
S_P[\tby]=\hf\int dtdt'[\tby(t)-\tby_{cl}(t)]\tG_{P,discr}^{-1}(t,t')[\tby(t')-\tby_{cl}(t')],
\ee
where
\be\label{clpointtr}
\tby_{cl}(t)=\tG_{\kappa g}(t)\tilde\sigma\frac1{1-\frac{N_a}{N_s}\tD\tilde\sigma\tG_{gg}\tilde\sigma}\tbx_{cl}
\ee
stands for the classical pointer trajectory $\by_{cl,\pm}=\by_{cl}$ with
\be\label{poincl}
\by_{cl}(t)=G^r_{\kappa g}(t)\frac1{1-\frac{N_a}{N_s}D^rG^r_{gg}}\bx_{cl}.
\ee
The inverse matrix, the~second factor on the right hand, acts on the vector $\bx_{cl}$ and represents the ``stroboscope physics''. The first factor, $G^r_{\kappa g}(t)$, embeds the result into the continuous time pointer dynamics. The~pointer propagator,
\vspace{6pt}
\be\label{pprop}
\tG_{P,discr}(t,t')=\frac{\tG_{\kappa^2}(t,t')}{N_a}+\tG_{g\kappa}(t)\frac1{N_s\tilde\sigma\tD^{-1}\tilde\sigma-N_a\tG_{gg}}\tG_{g\kappa}(t'),
\ee
is given in terms of the free propagator,
\be
\tG_{\kappa^2}(t,t')=\frac1{N_a}\sum_n\kappa_n^2\tG_n(t,t'),
\ee
and the average Green functions
\bea
\tG_{\kappa g,\ell}(t)&=&\frac1{N_a}\sum_n\kappa_ng_n\tG_n(t,t_\ell),\nn
\tG_{g^2,\ell,\ell'}&=&\frac1{N_a}\sum_ng^2_n\tG_n(t_\ell,t_{\ell'}).
\eea
The inverse on the right hand sides of \eq{clpointtr} and \eq{pprop} is taken for $2N_m-1\times2N_m-1$ matrices (indices are suppressed).

\end{document}